# Physical foundations and basic properties of magnetic skyrmions

**Alexei N. Bogdanov**[1, 2] **and Christos Panagopoulos**[3]

[1] IFW Dresden, Postfach 270116, D-01171 Dresden,
[2] Chirality Research Center, Hiroshima University, Higashi Hiroshima, Hiroshima 739-8526, Japan
[3] Division of Physics and Applied Physics, School of Physical and Mathematical Sciences, Nanyang Technological University, 21 Nanyang Link, Singapore 637371

**E-mail**: christos@ntu.edu.sg and a.bogdanov@ifw-dresden.de

**Abstract** Magnetic skyrmions (or vortices) are spatially inhomogeneous spin textures localized in nanoscale cylindrical regions. Topological protection and small size make skyrmions especially attractive for the study of spin topology and technologies wherein information is carried by the electron spin further to, or instead of the electron charge. Despite achievements in the synthesis of materials where axisymmetric magnetic skyrmions can be stabilized and characterized, there is disproportionate progress in elucidating the basic properties. The Perspective aims to bridge this gap and deliver an intelligible guide on the physical principles governing these magnetic whirls.

**Introduction**
Magnetic skyrmions were introduced theoretically in 1989 as a novel type of two-dimensional spatially localized states, stabilized in magnetic materials with broken inversion symmetry [1]. Typically, they emerge as right circular cylinders of axisymmetric spin texture (Fig. 1a) at magnetic fields *H* above saturation [1-4]. In Bloch-type magnetic skyrmions, for example, the magnetization vector *M* rotates with a fixed rotation sense in the planes perpendicular to the propagation vector *p*, namely, from an antiparallel direction at the axis to a parallel direction at distances far from the center (Fig. 1c). Bloch-type skyrmions stabilize in ferromagnets with $D_n$ symmetry [1, 4] and in the extended group of cubic helimagnets [3, 5-8]. This is just one of five possible skyrmion core configurations in non-centrosymmetric uniaxial ferromagnets (Box 2) [1, 4].

The study of magnetic skyrmions emerged from an apparent paradox. Mathematically, two- and three-dimensional localized structures are unstable in most condensed matter systems (the Hobart-Derrick theorem) [9]. Accordingly, localized structures such as magnetic skyrmions are not expected to exist. Hence, once induced they would rapidly collapse into linear singularities. However, magnetic materials with broken inversion symmetry do not obey the "prohibition" rule imposed by the Hobart-Derrick theorem [1]. In these low-symmetry systems, magnetic interactions imposed by the handedness of the underlying crystallographic structure (known as Dzyaloshinskii-Moriya interactions [10]) provide a physical mechanism, which prevents the collapse and stabilizes axisymmetric localized states with finite sizes [1, 11].

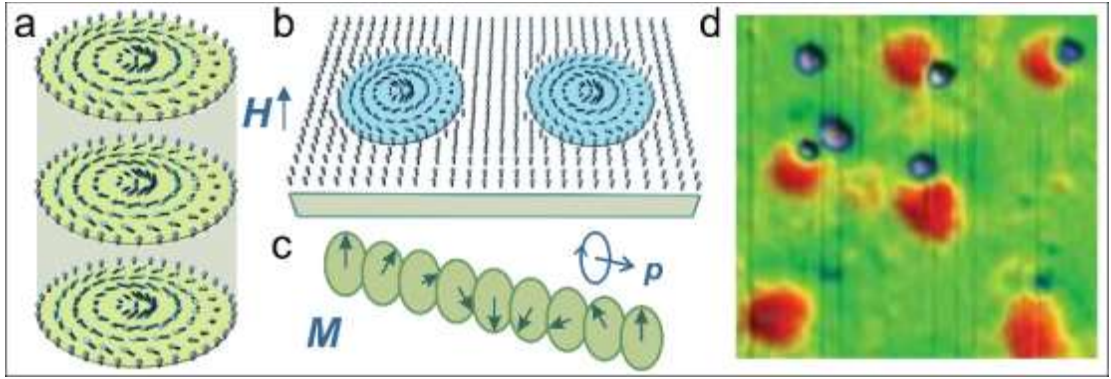

Figure 1| **a,b,c** | **Magnetic skyrmions** are axisymmetric nanoscale strings (a) emerging as ensembles of weakly repulsive "particles" at an applied magnetic field $H$ in homogeneously magnetized ferromagnets (b). Panels (a-c) depict most common (so called Bloch-type) skyrmion configurations with the magnetization rotating in the planes perpendicular to the propagation direction $p$ (c). **d** | Spin-polarized scanning tunneling microscopy images of isolated magnetic skyrmions with diameter approximately 6 nm (red spots) observed in a magnetically saturated FePd/Ir(111) nanolayer [13].

The experimental search for magnetic skyrmions commenced two decades after the theoretical prediction [5, 6]. These lead to the discovery of skyrmions in several groups of magnetic crystals, as well as in synthetic nanolayers and multilayers of magnetic metals [12-19]. Application of modern experimental methods such as Lorentz transmission electron microscopy, spin-polarized scanning tunneling microscopy and magneto-optical Kerr effect enabled imaging of skyrmions and their evolution (Fig. 1c) [13]. Resonant elastic x-ray scattering has been successful in determining the topological winding number of skyrmions, surface helicity angles of twisted skyrmions and even skyrmion rotation with well-defined dynamics in the presence of a magnetic field gradient [16]. Advances in characterization techniques and data analysis have encouraged combinatorial approaches. For example, complementary investigations now include neutron scattering, Lorentz transmission electron microscopy and high-field transport measurements to identify and study distinct topological spin textures [17].

The practicality of skyrmions indicated in the early theoretical studies "intrinsically stable localized magnetic inhomogeneities in the size of nanometers, which can be realized in low coercivity materials" [11], helped formulate a new paradigm in magnetic storage technologies. During the last decade, a variety of device concepts based on magnetic skyrmions emerged, providing basis for (*i*) highly mobile, (*ii*) low power, and (*iii*) super-dense magnetic data storage, and other spintronic applications [36, 41].

Subsequently, the field has been enriched with a plethora of experimental observations and results of numerical simulations. Despite impressive experimental accomplishments and encouraging application prospects, there has been relatively slow progress in understanding the physical principles in this rapidly extended field of science and technology. Although it is generally acknowledged that the mathematical formalism of magnetic skyrmions belongs to the physics of solitons, this domain of nonlinear physics is rather unfamiliar to modern multidisciplinary materials research. As a result, efforts for a coherent physical insight into the large number of data remain scarce.

In this Perspective, we initiate a discussion on the fundamental properties of magnetic skyrmions. Our primary focus is the emergence and evolution of these chiral spin textures in the well-studied non-centrosymmetric ferromagnets. We start from an analogy with shallow-water solitary waves (Box 1), discuss magnetic skyrmions as a special class of self-supporting localized states (solitons) and expound their stabilization mechanism. Using established mathematical tools of nonlinear physics, magnetic skyrmions are shown to exhibit properties expressed by methods of soliton physics (Box 2), and to emerge as axisymmetric strings with a fixed sense of rotation. Comparison with state-of-the-art experiments demonstrates the essential features of isolated skyrmions (Fig. 2) and skyrmion lattices (Fig. 3).

**Solitary waves and solitons**

Magnetic skyrmions (Fig. 1) belong to pattern formation phenomena widely spread in nature, manifesting as countable objects ("particles") in continuous fields. Such patterns are characterized by *spatial* and *temporal* dimensions. Tornados and typhoons, solitary surface ocean waves, and spherulites in chiral liquid crystals are just a few examples. In most physical systems, particle-like patterns emerge as unstable dynamic excitations, which gradually decay into a homogeneous state (*e.g.* tornados and typhoons). However, there exist a special class of "self-supporting" particle-like objects known as "*solitons*" [21-23].

Let us consider solitons as canonical case of travelling solitary waves on the water surface (Box 1). Commonly, once formed, such surface waves gradually spread and eventually decay. However, John Scott Russell (1834) realized an unusual stability of solitary surface waves travelling through shallow water channels. Following this observation, Joseph Boussinesq (1871) demonstrated theoretically that shallow water solitary waves are in fact stabilized by the seabed. These studies were developed further by Diederik Korteweg and Gustav de Vries (1898), who provided a detailed analysis of the localized and periodic solutions for shallow water waves (Box 1) [21].

The discovery of intrinsically stable, localized states introduced a new physical paradigm: a 'solid' particle-like object can be induced in continuous media and easily controlled and manipulated by external forces. This breakthrough was appreciated only decades later, in the second half of last century. The discovery of solitons (the term *"soliton"* was coined by Martin Kruskal and Norman Zabusky in 1969 [20]) in different domains of physics and the development of mathematical methods to describe their properties, formed one of the most important achievements in 20[th] century physics, the so-called "soliton revolution" [20, 21].

Box 1| **Shallow water solitons**

Mathematically, shallow water waves propagating along the *x*-axis are described by the nonlinear equation:

$$\frac{1}{v_0}\frac{\partial \eta}{\partial t} + \frac{\partial \eta}{\partial x} + \underbrace{\frac{h^2}{6}\frac{\partial^3 \eta}{\partial x^3}}_{\text{dispersion}} + \underbrace{\frac{3}{2h}\eta\frac{\partial \eta}{\partial x}}_{\substack{\text{interaction}\\ \text{with seabed}}} = 0, \qquad (B.1.1)$$

known as the *Korteweg-de Vries (KdV)* equation. Here the water surface elevation, $\eta(x, t)$ depends on $x$ and time $t$; $h$ is the water depth, $g$ is gravitation acceleration. The first two terms in the *KdV* equation yield solutions for non-dispersive waves propagating with speed $v_0 = \sqrt{gh}$; the next two terms describe the competing contributions of dispersion ($\propto h^2$) and the interaction with the seabed ($\propto 1/h$).

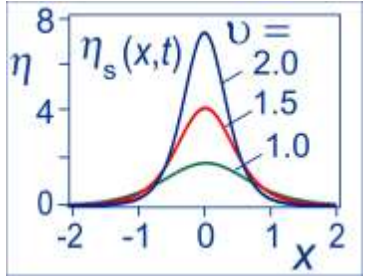

In extremely shallow water when $h$ is a factor 20 or more smaller than the wavelength $L$, the last term suppresses dispersion processes and stabilizes solutions for solitary waves with bell-shaped profiles (*solitons*) [21]

$$\eta_s(x,t) = \frac{2\,v^2}{\cosh^2[v(x - 4v^2 t - x_0)]}, (B.1.2)$$

where $v$ is a material parameter including the speed, width, and height of the wave. The calculated wave profiles $\eta_s(x, t)$ plotted for different values of parameter $v$ are in good agreement with laboratory reconstructions of Russell's soliton in a shallow water channel (see video in Related link).

**Skyrmion terminology**

The solution for a shallow water soliton (B.1.2) and the underlying physical mechanism stabilizing this state is a prototype for all solitonic states [23], including magnetic skyrmions. Originally, these solitonic states were introduced as "magnetic vortices" [1] and later renamed "magnetic skyrmions".

The term "skyrmion" was coined in early 1980's to designate solutions for three-dimensional solitons, for the field models [22] introduced by Tony Skyrme [53]. Subsequently, it has been applied to multidimensional non-singular localized states with nontrivial topology. In this extended meaning, "skyrmion" does not describe a fundamental entity with shared physical properties such as *electrons*, *protons*, or *quarks*. Instead, serves as an umbrella title for a variety of textures, heuristically attributed to topologically nontrivial, smooth localized structures in diverse areas of physics, from cosmology to condensed matter [55].

Following this trend, the term "skyrmion" has been used to designate magnetic localized states of diverse physical nature, such as magnetic bubble domains stabilized by demagnetization effects [25], solitonic states imposed by competing exchange interactions (frustration) [50] or Belavin-Polyakov instantons [4, 22].

Notably, magnetic skyrmions are not "topologically protected localized states". The nontrivial topology protects them from unwinding into saturated states but does not prevent them from collapsing into a linear singularity – a line with antiparallel magnetization – in the saturated phase. The presence of magnetic skyrmions with well-defined sizes is due to the stabilization supported by Dzyaloshinskii– Moriya interactions [1]. Further examples of solitonic states and associated stabilization mechanisms are discussed in the section on 'kinsmen' of magnetic skyrmions.

**Isolated magnetic skyrmions**
*Basic model.* Magnetically ordered states (ferromagnetic, antiferromagnetic *etc.*) are governed by quantum mechanical exchange coupling [24]. Nonetheless, many magnetically ordered materials can be described by classical field models with respect to the magnetization vector $M$ (*micromagnetics* [25]). Magnetic states in materials with broken spatial inversion symmetry can be described by basic model [10,11]

$$w(M) = A(\partial_i M)^2 + V(M) + w_D(M), \tag{1}$$

where $A(\partial_i M)^2 \equiv \sum_{i,j=1}^{3}(\partial M_j/\partial x_i)^2$ is the exchange energy with stiffness constant $A$, $V(M)$ is the "potential" term, including internal magnetic interactions and interactions with applied fields. For uniaxial ferromagnets $V(M) = K[1 - (M \cdot n)^2] + \mu_0[MH - M \cdot H]$ includes the uniaxial magnetocrystalline anisotropy with coefficient $K$ ($n$ is the unit vector along the high symmetry axis), and Zeeman energy. $\mu_0$ is vacuum permeability. In this model, the Dzyaloshinskii-Moriya energy density $w_D(M)$ is described by a combination of functionals linear in the first spatial derivatives of $M$ (so-called *Lifshitz invariants*):

$$\mathcal{L}_{ij}^{(k)} = M_i \partial M_j/\partial x_k - M_j \partial M_i/\partial x_k. \tag{2}$$

Energy contributions $\mathcal{L}_{ij}^{(k)}$ favor modulated states with magnetization rotation in the (*i*, *j*)-planes, propagating along the *k*-axis. They provide the stabilization mechanism for helical states [10] and magnetic skyrmions [1, 2]. For example, in non-centrosymmetric cubic ferromagnets, which belong to the crystallographic class *T* (MnSi, FeGe, Cu$_2$OSeO$_3$, and others) $w_D(M) = D\left(\mathcal{L}_{yx}^{(z)} + \mathcal{L}_{xz}^{(y)} + \mathcal{L}_{zy}^{(x)}\right) = D\, M \times \text{curl}(M)$ where D is the Dzyaloshinskii-Moriya constant [26]. These three Lifshitz invariants favor magnetization rotation in planes perpendicular to the propagation directions (as in *Bloch* domain walls [25]). Whereas, in uniaxial ferromagnets with $C_{nv}$ symmetry ($n = 2, 3, 6$), $w_D(M) = D\left(\mathcal{L}_{xz}^{(x)} - \mathcal{L}_{yz}^{(y)}\right)$. Here, the magnetization rotates along the propagation directions (as in *Néel* domain walls [25]).

The energy functional (1) includes interactions *essential* to stabilize magnetic skyrmions and neglects other energy contributions (such as demagnetization effects, magnetoelastic coupling, surface/interface induced interactions and instabilities, edge effects). The solutions derived within this simplified model [1-4, 11, 35] introduce fundamental properties of skyrmionic states in non-centrosymmetric ferromagnets. These findings offer the conceptual and mathematical basis for theoretical investigations of complex magnetic phenomena in magnetic nanolayers [28-32, 34] and artificial magnetic multilayers with intrinsic, or induced Dzyaloshinskii-Moriya interactions (see e.g. [37, 42]).

***Axisymmetric magnetic skyrmions.*** Solutions for isolated skyrmions $\theta(\rho), \psi(\phi)$ are shown in Box 2 and Figs. 2a, 2b). To date, three types of magnetic skyrmions have been identified in non-centrosymmetric ferromagnets. *Type-I* (*Bloch*-type) skyrmions have been observed in bulk samples and nanolayers of non-centrosymmetric cubic helimagnets [7, 8, 14, 19, 27-29]; *type-II* (*Neel*-type) discovered in rhombohedral ferromagnets $GaV_4S_8$ and $GaV_4Se_8$ with $C_{3v}$ symmetry belonging to a group of lacunar spinels [15], as well as in FePt/Ir nanolayers with surface-induced Dzyaloshinskii-Moriya interactions [4, 13, 30]; *type-III* reported in non-centrosymmetric Heusler alloys with $D_{2d}$ symmetry [19]. Solutions $\theta(\rho)$ (e.g. Fig. 2) are derived from equation (B.2.2) and described by the three characteristic parameters shown in (B.2.4) [4, 31].

A typical solution for magnetization profiles $\theta(\rho)$ is shown in Fig. 2b, together with experimental data for magnetic skyrmions in FePt/Ir nanolayers [4]. Skyrmion diameter, $L_s$ decreases with increasing applied magnetic field (Fig. 2b) [11, 31]. In contrast to magnetic bubble domains collapsing in moderate fields [25], skyrmions persist to high fields [11]. However, they lose radial stability and collapse when the applied magnetic field compresses them to sizes of few lattice constants [4, 32].

***Two scenarios of skyrmion evolution.*** At high magnetic fields, the energy of magnetic skyrmions $E$ (B.2.3) is positive and they exist as locally stable two-dimensional particles (cylinders) in a homogeneously saturated matrix (Figs. 1, 2). At lower fields, the skyrmion energy becomes negative. Hence, for $E$ (B.2.3) < 0 a spatially modulated phase composed of skyrmion "cells" (*e.g.* in the form of a hexagonal lattice (Fig. 3)) has lower energy than in the saturated state. Equation $E(H, K) = 0$ yields the critical field of the phase transition between the saturated state and a skyrmion lattice, $H_s(K)$ [2, 4]. In particular, for cubic helimagnets $H_s(0) = 0.801\ H_D$ (Fig.2b) [2]. The emergence of skyrmion lattices within the saturated phase, below $H_s(K)$, is possible only if isolated skyrmions can be easily nucleated. Otherwise, the saturated phase with embedded isolated skyrmions persists as a metastable state, far below $H_s$. From below, the presence of isolated magnetic skyrmions is restricted by the *elliptical* instability field, $H_{el}(K)$ [11]. Here, isolated skyrmions strip out abruptly into bands with one-dimensional helical modulations. In particular, for zero anisotropy $H_{el}(0) = 0.534\ H_D$ (Fig. 2c). Therefore, isolated skyrmions which arise as ensembles of weakly repulsive particles in the saturated phase, offer two scenarios of evolution with decreasing magnetic field. Either condense into spatially modulated phases (skyrmion lattices) below $H_s$ or remain as localized states in the metastable saturated phase and strip-out into one-dimensional helical states at $H_{el}$. Indeed, the spontaneous nucleation of skyrmion lattices with decreasing magnetic field has been observed in wedged nanolayers of cubic helimagnets. Here, magnetic skyrmions were nucleated at the edges and moved to the center of the sample [33, 34].

## Box 2| Solutions for magnetic skyrmions

For theoretical analysis of magnetic skyrmions, it's convenient to introduce spherical coordinates for the magnetization vector $M$ and cylindrical coordinates for the spatial variables $r$:

$$\boldsymbol{M}(\boldsymbol{r}) = M(\sin\theta\cos\psi, \sin\theta\sin\psi, \cos\theta), \quad \boldsymbol{r} = (\rho\cos\varphi, \rho\sin\varphi, z). \qquad (B.2.1)$$

Within the simplified model (1) the solutions for magnetic skyrmions are reduced to the form $\theta(\rho)$, $\psi(\varphi)$ (i.e. axisymmetric and homogeneous along the skyrmion axis, $z$) [1]. For non-centrosymmetric classes of uniaxial ferromagnets, solutions $\psi = \psi(\varphi)$ have been derived in analytical form [1]:

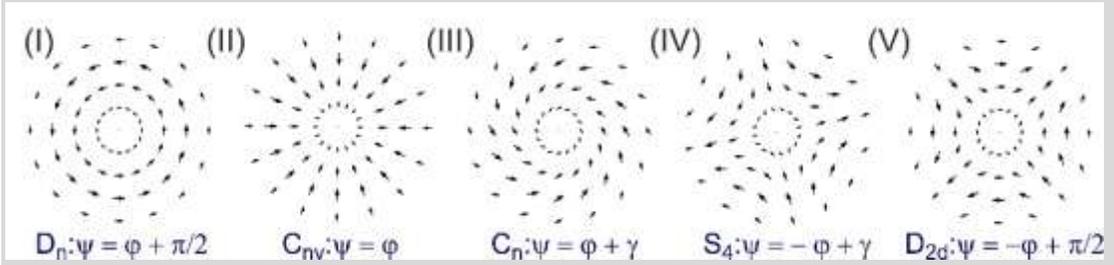

In-plane projections of the magnetization vector in this figure describe *five* different skyrmion core configurations [1-4]. In ferromagnets with $C_n$ and $S_4$ symmetries, $\gamma$ is a constant phase angle determined by the ratio of competing Dzyaloshinskii-Moriya interactions along orthogonal axes [1, 4]. For all five types of skyrmions, the equilibrium profiles $\theta(\rho)$ are derived from equation

$$A\left(\frac{d^2\theta}{d^2\rho} + \frac{1}{\rho}\frac{d\theta}{d\rho} - \frac{1}{\rho^2}\sin\theta\cos\theta\right) - \frac{D}{\rho}\sin^2\theta + f(\theta) = 0, \qquad (B.2.2)$$

with boundary conditions $\theta(0) = \pi$ and $\theta(\infty) = 0$. $A$ is the exchange stiffness constant, $D$ is the Dzyaloshinskii Moriya constant, and $f(\theta) = -K\sin\theta\cos\theta - \mu_0 MH\sin\theta$ is a "potential" term, including magnetic interactions independent of spatial gradients. The reduced energy

$$E = (2\pi)^{-1}\int_0^{2\pi}d\varphi\int_0^\infty w(\theta,\psi)\rho d\rho \qquad (B.2.3)$$

gives the difference between the magnetic skyrmion energy and the energy of the saturated state ($\theta = 0$). Model (B.2.1) introduces three characteristic parameters:

$$L_D = 4\pi A/|D|, \quad \mu_0 H_D = D^2 M/(2A), \quad K_0 = D^2/(4A), \qquad (B.2.4)$$

where $L_D$ is the period of helical modulations at zero field and anisotropy, $H_D$ is the saturation magnetic field and $K_0$ the critical value of uniaxial anisotropy (Fig. 2b, 2c).

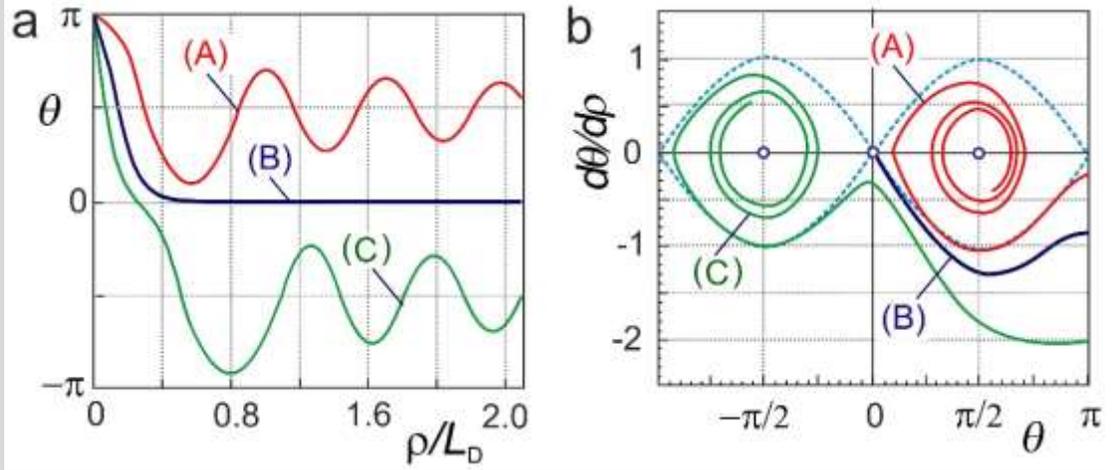

Plots (a) and (b) depict solutions of the initial value problem [$\theta(0) = \pi$, $d\theta/d\rho\,(0) = -\zeta$] for equation (B.2.2) with $H = 0$, $K = 2.8\,K_0$. Profiles $\theta_{(\zeta)}(\rho)$ (a) and the corresponding phase portraits ($\theta_{(\zeta)}$, $d\theta_{(\zeta)}/d\rho$) (b) are shown for different values of $\zeta$. In phase space ($\theta_{(\zeta)}$, $d\theta_{(\zeta)}/d\rho$) (b) localized solutions correspond to *separatrix* solutions (trajectory (B) in panel (b)). Localized solutions exist only for nonzero values of $D$. In centrosymmetric ferromagnets ($D = 0$), all phase trajectories end in one of the poles (for details see Refs. 4 and 35).

Spin-polarized scanning tunneling microscopy images in Fig. 2c demonstrate the evolution of weakly pinned isolated skyrmions at low fields in FePd/Ir nanolayers [4, 13, 30]. The formation of lattices at $H_s(T)$ is impeded by the restricted mobility of pinned, isolated skyrmions, which persist at their positions down to $H_{el}$. The interaction with neighboring skyrmions and the sample edges, hamper the strip-out transition at $H_{el}$. Instead, the elliptical distortions increase and skyrmions transform into a helical state [4].

The evolution of magnetic skyrmions in an applied field and the role of disorder in their stability and dynamics, encouraged intensive efforts on the synthesis of composite materials and the design of devices for prospective applications [36-42]. The technological aspect of magnetic skyrmions has been duly addressed in recent reviews by R. Wiesendanger [40] and A. Fert *et al.* [41]. In this Perspective, we focus on the fundamental properties and address the stabilization mechanism within the general principles of soliton physics.

*Magnetic skyrmions as solitons.* We may calculate the skyrmion energy $E$ (B. 2.3) with *ansatz*

$$\theta(\rho) = 4\arctan[\exp(-\rho/R)], \qquad (3)$$

where parameter $R$ represents the skyrmion core radius. This trial function is based on the well-known Landau-Lifshitz solution for isolated domain walls [25] and provides a good fit to the solutions of equation (B.2.2). The skyrmion energy $E$ (B. 2.3) calculated with ansatz (3) is reduced to the quadratic polynomial [4]

$$\mathcal{E}(R) = \mathcal{A} + \mathcal{B}R^2 - 3.02|D|R, \tag{4}$$

where $\mathcal{A} = 4.31 A$;  $\mathcal{B} = 1.59K + 1.39\mu_0 MH$. Minimization of $\mathcal{E}(R)$ yields the equilibrium skyrmion size

$$\bar{R}(H, K) = 1.51|D|/\mathcal{B}(H, K) \tag{5}$$

and the transition field $H_s(K)$. For cubic helimagnets, model (3) yields $H_s(0) = 0.760\, H_D$ (cf. with the rigorous value $0.801\, H_D$ (Fig. 2c)).

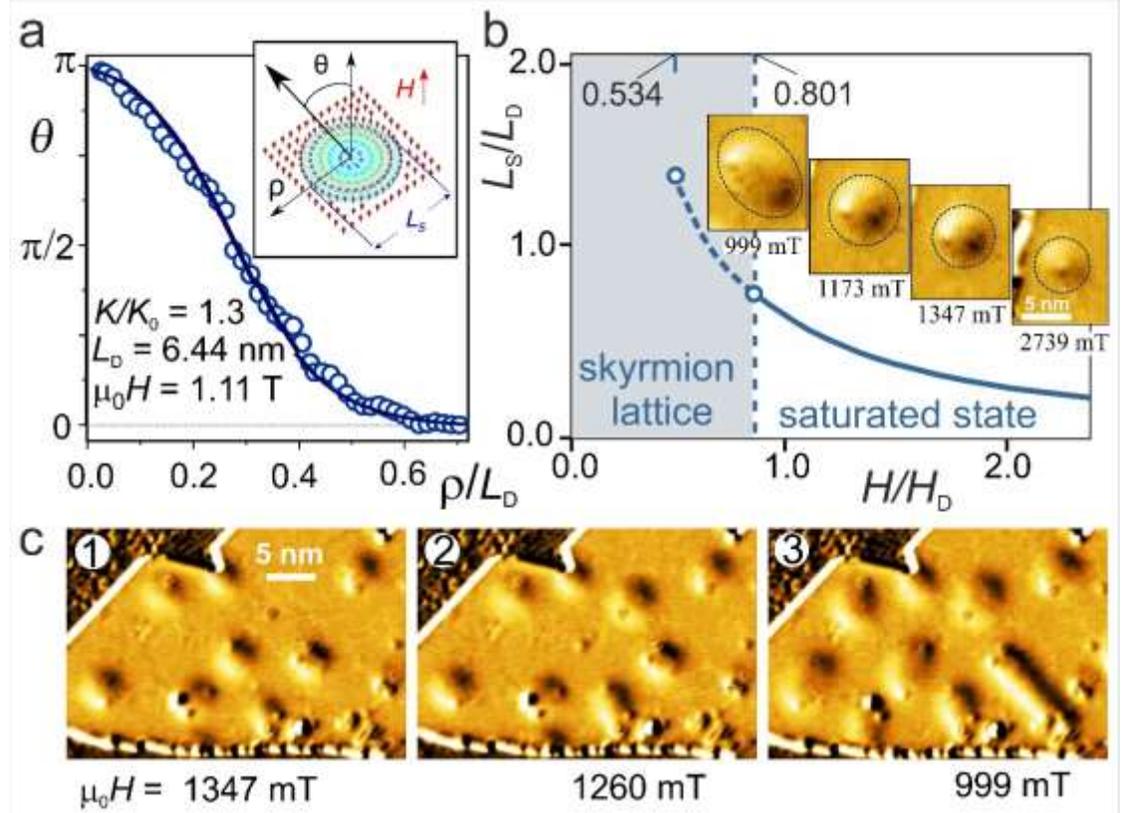

Figure 2| **Isolated skyrmions in the saturated state. a** | Magnetization profile $\theta\,(\rho/L_D)$ for an isolated skyrmion at high magnetic fields (circles are spin-polarized scanning tunneling microscopy data in a FePt/Ir (111) thin film [4]). The solid line is a fit to the data of the solution to equation (1) for $K/K_0 = 1.3$, $H/H_D = 0.32$. **b** | Equilibrium values of the isolated skyrmion size, $L_s$ as a function of applied magnetic field for zero uniaxial anisotropy [2]. Inset shows spin-polarized scanning tunneling microscopy images of an isolated skyrmion at different values of the applied magnetic field (adapted from [4]). Characteristic parameters $L_D$, $H_D$, and $K_0$ are introduced in Eq. B.2.4. **c** | Spin-polarized scanning tunneling microscopy images of magnetic skyrmions in a FePt/Ir (111) thin film for different values of applied magnetic field [4].

Energy $\mathcal{E}(R)$ includes two competing contributions: the Dzyaloshinskii-Moriya energy ($\propto -R$) favors unlimited extension of the skyrmion core ($R \to \infty$), while the energies of the uniaxial anisotropy and the interaction with the applied magnetic field ($\mathcal{B} \propto R^2$) tend to suppress the skyrmion ($R \to 0$). The equilibrium skyrmion size $\bar{R}$ (4) is formed due to the balance between counteracting intrinsic forces. Notably, similar competing processes (*dispersion* vs. *interaction with the seabed*) underlie the formation of shallow water solitons (Box 1). In both phenomena, the emerging localized states (*solitons*) correspond to the local minimum of the system, exhibiting a remarkable stability, protecting them against perturbations and preserving their shape. Interestingly, they demonstrate properties attributed to solitons and are described by methods of soliton physics (Box 2) (for details see Refs. 2, 4, and 35).

*Phase portraits of solutions.* Further insight into the nature of magnetic skyrmions can be obtained by solving Eq. (B.2.2) with initial values [$\theta(0) = \pi$, $d\theta/d\rho(0) = -\zeta$] (Box 2). Any localized solution of Eq. (B.2.2) (Fig. 2b) is among a set of parametrized profiles $\theta_{(\zeta)}(\rho)$ ($0 < \zeta < \infty$) and corresponds to a fixed value of $\zeta$. Typical profiles $\theta_{(\zeta)}(\rho)$ (a) and the corresponding phase portraits ($\theta_{(\zeta)}$, $d\theta_{(\zeta)}/d\rho$) (b) are plotted in Box 2. Most curves $\theta_{(\zeta)}(\rho)$ oscillate around lines $\theta_{1,2} = \pm \pi/2$ (a). The corresponding curves in phase ($\theta_{(\zeta)}$, $d\theta_{(\zeta)}/d\rho$) spiral around *attractor* points ($\pm \pi/2, 0$). In phase space ($\theta_{(\zeta)}$, $d\theta_{(\zeta)}/d\rho$), among the manifold of spiraling curves there is a singular curve ending at saddle point (0,0) (b). This so-called *separatrix* line corresponds to the localized solution of Eq. (B.2.2). As a matter of fact, localized solutions of Eq. (B.2.2) exist only for finite values of $D$. In centrosymmetric ferromagnets ($D = 0$), curves $\theta_{(\zeta)}(\rho)$ ($0 < \zeta < \infty$) oscillate around line $\theta_1 = \pi/2$ and the corresponding phase portrait curves spiral around attractor ($\pi/2, 0$). The geometrical images in a form of "shooting trajectories" and the phase portraits of the solutions (Box 2), demonstrate how Dzyaloshinskii-Moriya interactions stabilize solutions for localized states. The basic properties of magnetic skyrmions are elucidated without directly solving differential equation (B.2.2).

**Skyrmion lattices**

For a long time, homochiral long-period structures emerging in magnetic materials with broken inversion symmetry were observed in the form of one-dimensional modulations, *helices* [43]. Subsequently, it was established that in uniaxial non-centrosymmetric ferromagnets and nanolayers of cubic helimagnets, helical phases retain their thermodynamic stability only in weak magnetic fields. They transform by first-order processes into skyrmion lattices at a critical field, $H_1(T)$ [1-3, 31]. Above this field, magnetic skyrmion lattices (Fig. 3a) correspond to the global minimum of the system in a broad range of applied magnetic field and temperature [2, 31]. In experiment, the formation of hexagonal skyrmion lattices is commonly observed during the magnetic-field-induced phase transitions from the helical phase [8, 14, 15].

Solutions for skyrmion lattice cells describe a gradual localization of the skyrmion core and the expansion of the lattice period $L$ with the increase of an applied magnetic field (Fig. 3b) [2]. At $H_s(T)$, the lattice transforms into the saturated state by infinite expansion of the period ($L \to \infty$). Note that solutions for the skyrmion core have finite values at $H_s$. Hence, during transition into the saturated state the skyrmion lattice transforms into an ensemble of isolated skyrmions. Theoretically, this transition is reversible and on decreasing the applied magnetic field (below $H_s$) the ensemble of isolated skyrmions should re-condense to a skyrmion lattice. This is one of the possible scenarios for the evolution of a skyrmion texture.

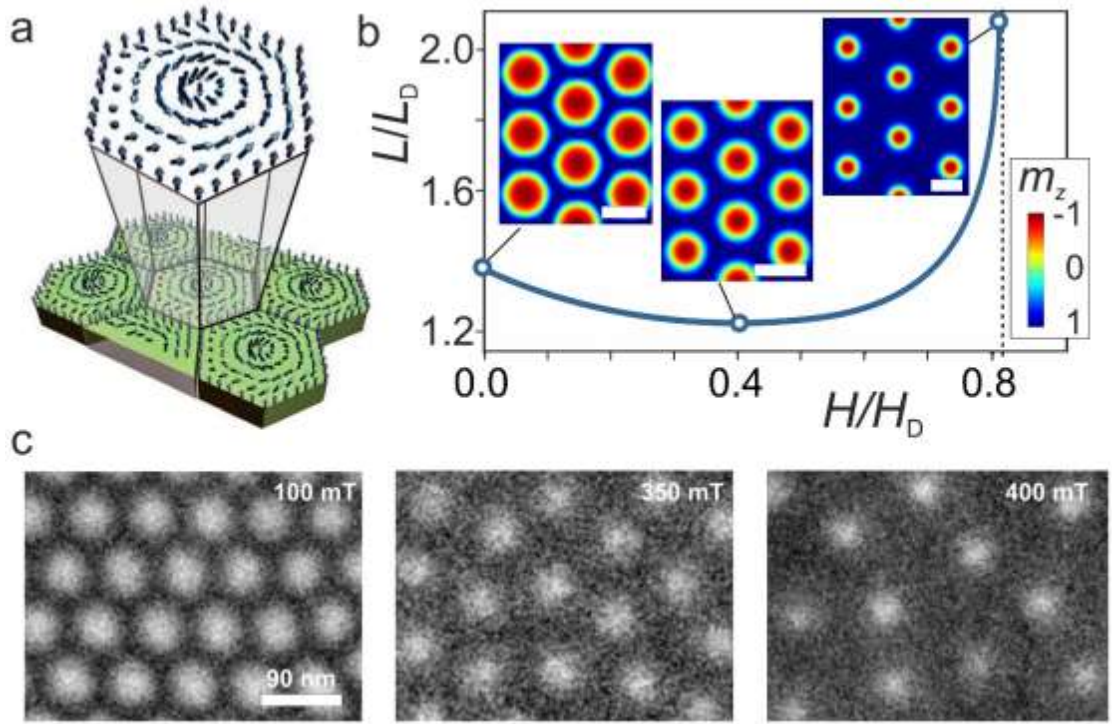

Figure 3| **Evolution of skyrmion lattices in an applied magnetic field. a**| Fragment of a hexagonal lattice with Bloch-type skyrmion cores. **b**| Equilibrium size of skyrmion cell $L$ as a function of applied magnetic field at zero anisotropy. Characteristic parameters $L_D$ and $H_D$ are introduced in Eq. B.2.4. Inset shows calculated contour plots $m_z(x,y)$ for $H/H_s = 0$ (1), 0.5 (2), 0.999 (3), $H_s = 0.801 H_D$. White bars indicate $L_D$ length. [28]. **c**| Images of skyrmion textures in a thin layer of FeGe for different applied magnetic fields (100 - 400 mT) recorded by off-axis electron holography at $T = 200$ K. Yellow lines indicate a dislocation causing distortion to the skyrmion cell. The internal structures of skyrmion cores are marked by squares (adapted from [29]).

The transformation of a skyrmion lattice to the saturated state develops continuously, like a second-order phase transition [44]. According to Landau theory, second-order transitions occur between a lower symmetry phase with a finite order parameter $\varsigma$ and a high symmetry phase ($\varsigma = 0$). During a phase transition, the order parameter gradually decreases and goes to zero at the transition point (*i.e.* the lower symmetry phase fades away at the transition point). Take for example, the order parameter of the ferromagnetic phase. Here, the magnetization modulus |***M***| goes to zero at the transition temperature and the system turns into the paramagnetic phase.

On the other hand, a skyrmion lattice does not vanish at the transition field. Instead, decomposes into the constituent elements (isolated skyrmions). According to a classification introduced by P.G. De Gennes, phase transitions with a continuous decomposition of lower symmetry phases belong to the *nucleation* type [45]. The continuous transition between helicoids and saturated phases proceeds in a similar way namely, through sublimation and re-sublimation of homochiral isolated domain walls [10]. Magnetic-field-driven transitions of bubble domain lattices and stripe domains into the homogeneous phase also belong to the nucleation type [25].

**Kinsmen of magnetic skyrmions**
From a mathematical perspective, solutions for multidimensional localized states arise only in field models containing either energy contributions linear with respect to spatial derivatives or with higher-order spatial derivatives [1, 10, 43]. The former is composed of Lifshitz invariants in eq. (2) and describe, for example, Dzyaloshinskii-Moriya interactions in magnetic compounds with broken inversion symmetry [1, 10, 46]. In condensed-matter there are no physical interactions underlying energy contributions with higher-order spatial derivatives. Interestingly however, a stabilization term, quartic in spatial derivatives (Skyrme mechanism), was introduced by Tony Skyrme, but to describe low-energy dynamics of mesons and baryons [22, 53]. Two-dimensional and three-dimensional localized solutions stabilized by the Skyrme mechanism are intensively investigated with so-called Faddeev-Skyrme models [22, 54]. The energy density functional

$$w(v) = a(\partial_i v)^2 + b(\partial_i v \times \partial_j v)^2 + V(v). \tag{6}$$

with the unity vector $v$ as the order parameter is representative for the family of field models [54] and consists of the common stiffness energy with constant $a$, the Skyrme energy with constant $b$, and the potential energy $V(v)$.

*Non-centrosymmetric systems.* In condensed matter systems with broken inversion symmetry such as ferroelectrics, chiral liquid crystals and mutliferroics, interactions analogous to Dzyaloshinskii-Moriya coupling can provide the stabilization mechanism for multidimensional solitons. Indeed, cholesterics and other chiral nematics host various textures with chiral modulations [46]. Among them, two-dimensional axisymmetric localized strings, analogous to magnetic skyrmions, have been observed in thin layers with strong perpendicular surface pinning (homeotropic anchoring) [47, 48]. Furthermore, axisymmetric skyrmions have been observed in layered oxides with spontaneous electric polarization (ferroelectrics) [49].

*Skyrmion-bubble hybrids***.** Localized magnetic patterns in nanolayers and multilayer magnetic architectures, emerge under the combined influence of surface/interface induced Dzyaloshinskii-Moriya interactions and magneto-dipolar coupling. These localized states may be considered `skyrmion-bubble hybrids', exhibiting magnetic properties attributed to chiral skyrmions and magnetic domains. They are distinct from the classical cylindrical domains (bubbles) observed in thin films with strong perpendicular magnetic anisotropy. Magnetic bubbles are areas of antiparallel magnetization, separated from the magnetically saturated matrix by thin domain walls [25]. Importantly, although magnetic bubbles and skyrmions have the same topology, they are fundamentally different physical objects. Bubbles are intrinsically unstable magnetic domains, which can be stabilized by surface demagnetization effects in confined magnets and depend on the shape of a magnetized body [25].

Skyrmion-bubble hybrids are a new class of chiral localized states. The modulation of the geometry or composition of a magnetic multilayer, promises fine control over the frequency response associated with skyrmions and the corresponding gyration of the topological charge, which is dependent on the inter-layer dipolar coupling within the heterostructure. For example, raising this coupling may increase the skyrmion radius and tune the frequency and field response of its resonance. This intrinsic adjustability highlights one of the potential advantages of chiral magnetic multilayers.

*Exchange modulations.* These modulated textures emerge in an extended group of centrosymmetric magnetic compounds and are stabilized by competing non-local exchange interactions [43]. Unlike homochiral and long-periodic modulations induced by Dzyaloshinskii-Moriya interactions, exchange modulations are characterized by periods of a few lattice constants and arbitrary rotation sense. These give rise to a numerous short period spirals in rare earth metals and related materials [43]. There has been evidence for nanoscale magnetic heterostructures exhibiting multidimensional short period modulated states supported by complex non-local exchange interactions [12]. Instead of continuum (field) models, such short period modulations are described either within discrete models [4, 32, 50, 51] or by applying methods of quantum magnetism [12, 52]. The fundamental properties of magnetic skyrmions (including stability criteria) have been derived within field models and do not apply to properties of short period modulations (independent of the stabilization mechanism). Thus, the multidimensional short period modulated states [12, 32, 50, 51] should not be confused with magnetic skyrmions and other solitons. In a continuum limit, discrete models for exchange modulations reduce to energy functionals of the type in eq. (5) [43, 56, 57] and establish relations between exchange modulations and fields models, with the Skyrme stabilization mechanism.

**Summary and outlook**

This Perspective addresses the physical foundations of magnetic skyrmions, a rapidly developing field in science and technology. It aims to serve as an intelligible guide on the essential principles governing the properties of magnetic skyrmions in condensed matter physics. Conceptualization of the formation and evolution of these solitonic states helps bring theoretical findings to the laboratory bench. Specifically, the underlying crystallographic structure of chiral magnets with broken inversion symmetry induces competing magnetic interactions, with the emerging localized states exhibiting remarkable stability, protecting them against perturbations and preserving their shape. Practically, the basic physical properties illustrate that these spin textures can be created and manipulated using readily available tools in a plethora of bulk crystals and synthetic architectures.

**Acknowledgements:** The authors acknowledge A. Fert and R. Wiesendanger for insightful discussions. A.N.B. thanks M. Ochi and K. Inoue for hospitality and collaboration during his stay at Hiroshima University. This work was supported in Germany by the Deutscher Forschungs-gemeinschaft through SPP2137 "Skyrmionics",

**Key references:**

1. Bogdanov, A. N. & Yablonskii, D. A. Thermodynamically stable "vortices" in magnetically ordered crystals. The mixed state of magnets. *Sov. Phys. JETP*. **68**, 101-103 (1989).
2. Bogdanov, A. & Hubert, A. Thermodynamically stable magnetic vortex states in magnetic crystals. *JMMM* **138**, 255-269 (1994).
3. Rößler, U. K., Bogdanov, A. N. & Pfleiderer C. Spontaneous skyrmion ground states in magnetic metals. *Nature* **442**, 797-801 (2006).
4. Leonov, A. O. *et al.* The properties of isolated chiral skyrmions in thin magnetic films. *New Journal of Physics* **18,** 065003 (2016).
5. Mühlbauer, S. *et al*. Skyrmion lattice in a chiral magnet. *Science* **323**, 915-919 (2009).
6. Pappas C., et al., Chiral paramagnetic skyrmion-like phase in MnSi, *Phys. Rev. Lett.* **102**, 197202 (2009).
7. Yu, X. Z. *et al*. Real-space observation of a two-dimensional skyrmion crystal. *Nature*, **465**, 901-903 (2010).
8. Yu, X. Z. *et al*. Near room-temperature formation of a skyrmion crystal in thin-films of the helimagnet FeGe. *Nat. Mater.* **10**, 106–109 (2011)
9. Derrick, G. H. Comments on nonlinear wave equations as models for elementary particles. *Journal of Mathematical Physics* **5,** 1252–1254 (1964).
10. Dzyaloshinskii, I. E. Theory of helicoidal structures in antiferromagnets. *Sov. Phys. JETP* **19**, 960–971 (1964).
11. Bogdanov, A. & Hubert, A. The properties of isolated magnetic vortices. *Phys. Stat. Sol. (b)* **186**, 527 (1994).
12. Heinze, S. *et. al.* Spontaneous atomic-scale magnetic skyrmion lattice in two dimensions. *Nat. Phys.* **7**, 713–718 (2011).
13. Romming, N. *et. al.* Writing and deleting single magnetic skyrmions. *Science* **341,** 636-639 (2013).
14. Milde, P. *et al*. Unwinding of a skyrmion lattice by magnetic monopoles. *Science* **340**, 1076–1080 (2013).
15. Kézsmárki, I. *et al*. Néel-type skyrmion lattice with confined orientation in the polar magnetic semiconductor GaV$_4$S$_8$. *Nat. Mater.* **14**, 1116–22 (2015).
16. Zhang, S.L., Wang, W.W., Burn, D.M. et al. Manipulation of skyrmion motion by magnetic field gradients, *Nat Commun* **9,** 2115 (2018).
17. Fujishiro, Y., Kanazawa, N., Nakajima, T. et al. Topological transitions among skyrmion- and hedgehog-lattice states in cubic chiral magnets, *Nat Commun* **10,** 1059 (2019).
18. Zefang D. et al. Observation of Magnetic skyrmion bubbles in a van der Waals Ferromagnet Fe$_3$GeTe$_2$ *Nano Lett.* **20**, 868-873 (2020).
19. Nayak, A. K. *et al.* Magnetic antiskyrmions above room temperature in tetragonal Heusler materials. *Nature* **548**, 561-566 (2017).
20. Zabusky, N. J. & Kruskal, M. D., Interaction of solitons in a collisionless plasma and the recurrence of initial states*, Phys. Rev. Lett.* **15**, 240—243 (1965).
21. Remoissenet, M.: *Waves Called Solitons. Concepts and Experiments*. Springer-Verlag, 2003. — 328 p.
22. Manton, N. and Sutcliffe, P. *Topological Solitons* (Cambridge: Cambridge University Press) (2004).
23. Bogdanov A. N. & Panagopoulos C. The emergence of magnetic skyrmions. *Physics Today* **73**, 44-49 (2020) (for extended version see arXiv: 2003.09836).
24. White R. M.: *Quantum Theory of Magnetism.* Springer-Verlag Berlin Heidelberg 2007. — 362 p.
25. Hubert, A., & R. Schäfer, *Magnetic Domains* (Springer, Berlin) (1998).
26. Bak P. & Jensen M. H. Theory of helical magnetic structures and phase transitions in MnSi and FeGe. *J. Phys. C: Solid State Phys.* **13** L881-L885 (1980)
27. White, J. S. *et al*. Electric-field-induced skyrmion distortion and giant lattice rotation in the magnetoelectric insulator Cu$_2$OSeO$_3$. *Phys. Rev. Lett.* **113**, 107203 (2014).
28. McGrouther, D. *et al.* Internal structure of hexagonal skyrmion lattices in cubic helimagnets. *New J. of Phys.* **18,** 095004 (2016).
29. Kovács, A. *et al.* Mapping the magnetization fine structure of a lattice of Bloch-type skyrmions in an FeGe thin film. *Appl. Phys. Lett.* **111,** 192410 (2017)
30. Romming, N., Kubetzka, A., Hanneken, C., Bergmann, K. V. & Wiesendanger, R. Field-dependent size and shape of single magnetic skyrmions. *Phys. Rev. Lett.* **114,** 177203 (2015).
31. Wilson, M. N., Butenko, A. B., Bogdanov, A. N. & Monchesky T. L. Chiral skyrmions in cubic helimagnet films: The role of uniaxial anisotropy. *Phys. Rev. B* **89,** 094411 (2014).
32. Siemens, A., Zhang, Y., Hagemeister, J., Vedmedenko, E. Y. & R Wiesendanger, R. Minimal radius of magnetic skyrmions: statics and dynamics. *New J. of Phys.* **18,** 045021 (2016).
33. Yu, X. Z. *et al*. Variation of skyrmion forms and their stability in MnSi thin plates. *Phys. Rev. B* **91**, 054411 (2015).
34. Leonov, A. O., et *al.* Chiral Surface Twists and Skyrmion Stability in Nanolayers of Cubic Helimagnets. *Phys. Rev. Lett.* **117,** 087202 (2016).
35. Bogdanov, A. & Hubert, A. The stability of vortex-like structures in uniaxial ferromagnets. *JMMM* **195**, 182-192 (1999).
36. Fert, A., Cros, V. & Sampaio, J. Skyrmions on the track. *Nat. Nanotechnol.* 8, 152-156 (2013).
37. Moreau-Luchaire C. et al., Additive interfacial chiral interaction in multilayers for stabilization of small individual skyrmions at room temperature, *Nature nanotechnology* **11**, 444-448 (2016).
38. Soumyanarayanan, A., Reyren, N., Fert, A. & Panagopoulos, C. Emergent phenomena induced by spin–orbit coupling at surfaces and interfaces. *Nature* **539**, 509-517 (2016).
39. Woo S. *et al*. Observation of room temperature magnetic skyrmions and their current-driven dynamics in ultrathin Co films, *Nature Materials* **15**, 501- 506 (2016).
40. Wiesendanger, R. Nanoscale magnetic skyrmions in metallic films and multilayers: a new twist for spintronics. *Nat. Rev. Materials* **1**, 1-11 (2016).
41. Fert, A., Reyren, N. & Cros, V. Magnetic skyrmions: advances in physics and potential applications. *Nat. Rev. Materials* **2**, 1-15 (2017).
42. Duong, N.K., Raju, M., Petrović, A.P., Tomasello, R., Finocchio, G. & Panagopoulos, C., Stabilizing zero-field skyrmions in Ir/Fe/Co/Pt thin film multilayers by magnetic history control, *Appl. Phys. Lett.* **114**, 072401 (2019).

43. Izyumov, Yu. A. Modulated, or long-periodic, magnetic structures of crystals, *Sov. Phys. Usp.* **27,** 845-867 (1984).
44. Landau, L. D. & Lifshitz, E. M. *Statistical physics*. 3th Edition, Part 1. (Pergamon Press, Oxford, 1980).
45. De Gennes P. G. Fluctuations, Instabilities, and Phase transitions (NATO ASI Ser. B vol 2) ed T Riste (Plenum, New York, 1975)
46. Wright D. C. & Mermin N. D., Crystalline liquids: the blue phases, Rev. Mod. Phys. **61**, 385-432 (1989).
47. Leonov, A. O *et al.* Theory of skyrmion states in liquid crystals, *Phys. Rev. E* **90**, 042502 (2014).
48. Ackerman, P. J. *et al.*, Two-dimensional skyrmions and other solitonic structures in confinement-frustrated chiral nematics*, Phys. Rev. E* **90**, 012505 (2014).
49. S. Das *et al.*, Observation of room-temperature polar skyrmions, *Nature* **568**, 368-372 (2019).
50. Okubo, T., Chung, S. & Kawamura, H. Multiple-q states and the skyrmion lattice of the triangular-lattice Heisenberg antiferromagnet under magnetic fields. *Phys. Rev. Lett*. **108**, 017206 (2012).
51. Leonov A. O. & Mostovoy M. Multiply periodic states and isolated skyrmions in an anisotropic frustrated magnet. *Nat. Comm.* **6**, 8275 (2015).
52. Romming N. *et al.* Competition of Dzyaloshinskii-Moriya and higher-order exchange interactions in Rh/Fe atomic bilayers on Ir (111). *Phys. Rev. Lett.* **120**, 207201 (2018).
53. Skyrme, T. H. A non-linear field theory. *Proc. R. Soc. A* **260**, 127–138 (1961).
54. Faddeev, L. D. Some comments on the many-dimensional solitons. Lett. Math. Phys. **1**, 289–293 (1976).
55. Brown, G. E. & Rho, M. (ed) *The Multifaceted Skyrmion* (Singapore: World Scientific) (2010).
56. Hubert, A. *Theorie der Domänenwände in geordneten Medien*, (Springer, Berlin) (1974).
57. Melnichuk, P. I., Bogdanov, A. N., Rößler, U. K., & Müller, K.-H. Hubert model for modulated states in systems with competing exchange interactions. *JMMM* **248**, 142-150 (2002).

**RELATED LINKS**
Pizzo, N., "Shallow water wave generation," www.youtube.com/watch?v=w-oDnvbV8mY.